\documentclass{article}
\usepackage[T1]{fontenc}
\usepackage[utf8]{inputenc}
\usepackage[english]{babel}
\usepackage{graphicx} 
\usepackage{amsmath}
\usepackage{amssymb}
\usepackage{latexsym}
\usepackage{cancel}
\usepackage{braket}
\usepackage{bm}
\usepackage{siunitx}
\usepackage{multirow}
\usepackage{booktabs}
\usepackage{xcolor}
\usepackage{verbatim}
\usepackage{subfiles}
\usepackage{subfigure}
\usepackage[backend=bibtex,style=chem-acs]{biblatex}
\usepackage{hyperref}
\usepackage{csquotes}
\usepackage{tikz}
\usepackage{geometry}
\usepackage{setspace}
\usepackage{simplewick}
\usepackage{algorithm}
\usepackage{algpseudocode}
\usepackage{listings}
\usepackage{mathtools}
\usepackage{authblk}

\usetikzlibrary{shapes,
                tikzmark}
\tikzset{every tikzmarknode/.style={%
        draw=red, semithick, inner sep=2pt}
        }

\usetikzlibrary{shapes.geometric,
                arrows}
\usetikzlibrary{tikzmark, positioning, fit, shapes.misc}
\usetikzlibrary{decorations.pathreplacing, calc}

\tikzset{brace/.style={decorate, decoration={brace}},
    brace mirrored/.style={decorate,decoration={brace,mirror}},
}
                
\tikzstyle{startstop} = [rectangle, rounded corners, 
minimum width=1cm, 
minimum height=1cm,
text centered, 
draw=black, 
fill=green!30]

\tikzstyle{process} = [rectangle, 
minimum width=1cm, 
minimum height=1cm, 
text centered, 
text width=4cm, 
draw=black]

\tikzstyle{decision} = [diamond, 
minimum width=0.5cm, 
minimum height=0.5cm, 
aspect=2,
text centered,
text width=3cm, 
draw=black, 
fill=pink!30]

\tikzstyle{arrow} = [thick,->,>=stealth]

\definecolor{codegreen}{rgb}{0,0.6,0}
\definecolor{codeblue}{rgb}{0,0,1}
\lstdefinestyle{mystyle}{commentstyle=\color{codegreen},
                         keywordstyle=\color{codeblue},
                         stringstyle=\color{magenta},
                         basicstyle=\ttfamily\footnotesize,
                         language=[90]Fortran,
                         breakatwhitespace=false,         
                         breaklines=true,                 
                         captionpos=b,                    
                         keepspaces=true,                                   
                         showspaces=false,                
                         showstringspaces=false,
                         showtabs=false,                  
                         tabsize=2}
\title{A novel implementation of CCSD analytic gradients using Cholesky decomposition of the two-electron integrals and Abelian point-group symmetry}
\author[1]{Luca Melega}
\author[1]{Tommaso Nottoli}
\author[2]{J\"urgen Gauss}
\author[1]{Filippo Lipparini}
\affil[1]{Dipartimento di Chimica e Chimica Industriale, University of Pisa, Via G. Moruzzi 13, 56124, Pisa, Italy}
\affil[2]{Department Chemie, Johannes Gutenberg-Universit\"at Mainz, Duesbergweg 10-14, 55128, Mainz, Germany}
\date{}

\begin{document}

\maketitle

%
%

\section*{Abstract}

We present a novel and efficient implementation of coupled-cluster with singles and doubles (CCSD) analytic gradients that combines the Cholesky decomposition (CD) of electron-repulsion integrals with the exploitation of Abelian point-group symmetry.  This approach is particularly effective for medium-sized and large symmetric molecular systems. The CD of two-electron integrals is performed using a symmetry-adapted two-step algorithm, while the derivatives of the Cholesky vectors are computed with respect to symmetry-adapted nuclear displacements and contracted on-the-fly with the CCSD density matrices. 
Geometry optimizations of symmetric systems with several hundreds of basis functions have been carried out to assess the efficiency of our implementation and to quantify the computational gain provided by the exploitation of point-group symmetry.

\section{Introduction}

The accurate prediction of molecular properties for chemically relevant systems has long been a central goal of quantum chemistry~\cite{helgaker_properties}. For instance, geometrical gradients are necessary for locating local minima on potential energy surfaces~\cite{Pulay1977grad}, determining thus equilibrium structures and properties, as well as for finding transition states~\cite{schlegel2011geoopt}. Similarly, computed electric and magnetic properties, as well as force constants, can be used for the simulation of various spectroscopies~\cite{helgaker1999nmr,gauss2002electron,pulay2014analytical}.

Coupled-cluster (CC) methods~\cite{Shavitt_Bartlett_2009} are widely regarded as the gold standard for accurate energy and property calculations in systems where static correlation is negligible, thanks to their inherent accuracy, size-extensivity, and systematic improvability. Among these, the CCSD approach~\cite{purvis1982ccsd}, which includes all connected single and double excitations from the reference wavefunction, offers a good balance between accuracy and computational cost. Nevertheless, its formal $\mathcal{O}(N^6)$ scaling and substantial memory requirements restrict routine applications to medium-sized molecules, typically those with up to 10–15 heavy (e.g., non-hydrogen) atoms.

The development of analytic CC gradients has been an active area of research since the 1980s~\cite{adamowicz1984lambda,salter1989analytic}. Early implementations were hindered by the non-variational nature of CC theory, which appeared to require the evaluation of perturbed wavefunction parameters (amplitudes and molecular orbital coefficients). However, Adamowicz, Laidig, and Bartlett~\cite{adamowicz1984lambda} demonstrated, using the interchange theorem of perturbation theory~\cite{dalgarno1958perturbation}, that this step can be avoided by solving an additional perturbation-independent system of linear equations—the so-called Lambda equations. Salter, Trucks, and Bartlett~\cite{salter1989analytic} later extended the theory up to the CC singles, doubles, and triples (CCSDT) level. The first working implementation was reported more or less at the same time by Scheiner \textit{et al.}~\cite{scheiner1987analytic} for the special case of closed-shell CCSD. The extension to open-shell CC treatment was pushed forward in the group of Bartlett in 1991~\cite{gauss1991coupled,GAUSS1991207rohf,gauss1991qrhf}. CC gradients with triples were implemented by several authors for CCSD with a perturbative treatment of triple excitations (CCSD(T))~\cite{scuseria1991analytic,lee1991analytic,WATTS19921open,watts93}, iterative approximations to CCSDT  \cite{Scuseria88, Gauss00}, and finally in 2002 for the full CCSDT model~\cite{gauss2002ccsdt}. A general CC gradient implementation has been presented in 2003 by Kállay, Gauss, and Szalay~\cite{kallay2003analytic}.
A major conceptual advance was the introduction of the Lagrangian approach by Helgaker and co-workers~\cite{helgaker1982simple,koch1990lagrangian}, which greatly simplified the derivation of gradient expressions and eliminated the need for the interchange theorem. This approach has since been widely adopted~\cite{hald2003lagrangian,feng2019implementation,bozkaya2016analytic,schnack2022efficient}.

To alleviate the high computational cost of CC methods, numerous techniques have been 
developed~\cite{hampel1996local,schuetz2001,werner2011efficient,neese2009efficient,neese2009efficient2,riplinger2013dlpno,wei2009local,rolik2011general,kjaergaard2017divide,epifanovsky2013general,Nottoli2023,peng2017low,parrish2012hyper,parrish2019rank,hohenstein2022rank}
and used by quantum chemists over the years in order to improve the efficiency of CC calculations.
Among these, approaches that aim to reduce the scaling of CC calculations by exploiting the local nature of dynamic electron correlation have seen recently a large rise in popularity~\cite{hampel1996local,schuetz2001,werner2011efficient,riplinger2013triple,wei2009local,rolik2011general,ma2018explicitly}. To be mentioned here are in particular the local CC (LCC) method by Werner and Sch\"utz~\cite{schuetz2001,werner2011efficient} and the domain-based local pair-natural orbital CC (DLPNO-CC) method developed by Neese and co-workers~\cite{neese2009efficient2,riplinger2013dlpno,guo2018dlpno}; both
feature linear scaling of the cost with respect to the system size. However, these approaches rely on orbital localization and multiple thresholds to enforce the locality of correlation, factors that complicate the derivation of analytic gradient expressions and seriously hamper their implementation~\cite{rauhut2001analytical,datta2016analytic}.
On the other hand, rank-reducing strategies have been developed that lower the computational cost of quantum-chemical methods, despite keeping their overall scaling~\cite{deprince2013accuracy,epifanovsky2013general,Nottoli2023}. These schemes often use low-rank approximations of the involved tensors such as the electron-repulsion integral (ERI) matrix in order to drastically reduce computational cost and memory 
requirements. They also often allow implementations that can be easily parallelized, vectorized, and rewritten through highly optimized matrix-matrix products. The resolution-of-the-identity (RI)/density fitting (DF)~\cite{Whitten1973,Dunlap79,Vahtras93,Feyereisen1993,Eichkorn95,Weigend97,Weigend2002,Sierka03,Sodt2006}
and Cholesky decomposition (CD)~\cite{beebelinderberg1977cd,Roeggen1986,Koch2003,Roeggen2008,Weigend2009,Pedersen2024cd} approximations both belong to this class of methods. In RI/DF, the four-centers ERIs are rewritten in terms of three-centers intermediates by expanding product densities through the introduction of an auxiliary basis set consisting of pre-optimized functions. The CD of two-electron integrals, first suggested by Beebe and Linderberg in 1977~\cite{beebelinderberg1977cd}, exploits the rank-deficiency of the ERI matrix to yield a representation in terms of Cholesky vectors, which enables an efficient compression of the information stored in the full tensor. 

The formulation and implementation of analytic gradients is for schemes that employ RI/DF or CD considerably simpler than for those that exploit the locality of electron correlation~\cite{bostrom2013analytical,bozkaya2014derivation,aquilante2008analytic,bostrom2014analytical,delcey2014analytical}. Consequently, gradients for CD-based CC schemes have been reported~\cite{feng2019implementation,schnack2022efficient,bozkaya2016analytic,bozkaya2017triples}.
Feng \textit{et al.}~\cite{feng2019implementation} reported an implementation for CCSD as well as equation-of-motion CCSD (EOM-CCSD). While it is possible to perform large-scale computations with their implementation (within the Q-Chem program package~\cite{qchem5}), the need to compute and store all perturbed Cholesky vectors renders this implementation not optimal. The gradient implementation by Schnack-Petersen \textit{et al.}~\cite{schnack2022efficient} resolves this issue by exploiting the analogy between CD and RI/DF. The reported sample calculations demonstrate the efficiency of their implementation within the $e^T$ package~\cite{eT1.0}, but one should note that their implementation does not exploit point-group symmetry. Although symmetry is less relevant for very large molecules (which typically lack symmetry), it provides a significant advantage for medium-sized systems—the typical application range of CC theory. Moreover, symmetry exploitation becomes particularly important because RI/DF and CD do not alter the formal scaling of CC computations, which remain computationally demanding even with these approximations. 

In light of this, we present a new implementation of CCSD analytic energy gradients based on the CD of the ERI tensor that also exploits Abelian point-group symmetry. Our choice of CD over RI/DF is consistent with the developments referenced in the previous paragraph and is due to the fact that the accuracy of the latter is limited by the choice of the specific fitted auxiliary basis and cannot be rigorously and systematically controlled. In contrast, the accuracy of CD is rigorously determined by the threshold that truncates the decomposition, which is set $\textit{a priori}$ and is the only user-defined parameter in the procedure. This makes CD especially desirable when coupled with highly accurate methods. Our code has been incorporated in a development version of the {\sc CFOUR} suite of programs~\cite{cfour,code_cfour}.

We start in the following (section~\ref{section:theory}) by recounting the benefits of CD in quantum chemistry with a focus on CD for the derivatives of ERIs. This is followed by a discussion of the analytic expression for CCSD gradients, starting from the definition of a CC Lagrangian, which in turn leads to the Lambda equations that need to be solved to compute CC derivatives (section~\ref{section:theory}). Relevant computational details concerning our implementation are given, with a focus on the treatment of differentiated two-electron integrals in the Cholesky representation (also in section~\ref{section:theory}), and the explicit inclusion of Abelian point-group symmetry within our implementation (section~\ref{section:comput}). In section~\ref{section:test} we present numerical results concerning timings of CD-CCSD geometry optimizations of symmetric systems and the efficiency of OpenMP parallelization and symmetry-adaptation within our code. Finally, we provide a summary and an outlook on future work.

\section{Theory}

\label{section:theory}

In this section, we outline the theoretical foundations behind the CD for the two-electron integrals and the CD for their derivatives as well as
provide a brief summary on the derivation of the relevant expressions for CCSD analytic gradients.

\subsection{Cholesky decomposition of two-electron integrals}

\label{subsection:cd}

Since the ERI matrix (in Mulliken notation) is symmetric and positive semidefinite, it can be represented via a Cholesky decomposition in the following way:
\begin{equation}
    (\mu \nu \vert \rho \sigma) \approx \sum_P L_{\mu \nu}^P L_{\rho \sigma}^P ,
\end{equation}
where $L_{\mu \nu}^P$ refers to the $\mu \nu$ element of the $P$th Cholesky vector (CV). Since the tensor itself is positive semidefinite, its CD is not unique.

The number of two-electron integrals formally scales as $\mathcal{O}(N^4)$, with $N$ being the number of basis functions for the considered system. However, the ERI tensor is not full-rank, given that the number of its non-zero eigenvalues only scales linearly with the size of the basis set. Since the effect of CD is to remove linear dependencies between the columns (and rows) of a matrix, thus eliminating zero or near-zero eigenvalues, the actual number of CVs that need to be computed (which we will call $N_{ch}$) scales itself as $\mathcal{O}(N)$. As a consequence, $N_{ch} << \tfrac{N(N+1)}{2}$, that is, the number of CVs required to numerically represent the integrals is significantly smaller than the number of all unique basis pairs $\vert \rho \sigma)$, leading to reduced 
RAM (Random-Access Memory) requirements.

As the derivatives of the ERIs no longer constitute a positive semidefinite matrix, they cannot be directly decomposed via a CD. However, it is possible to derive CD-type expressions for them via differentiation of the CD expressions for the undifferentiated ERIs~\cite{feng2019implementation,giao-mp2, Gauss23, Burger25} or, alternatively, by exploiting the analogy of RI/DF and CD. The obtained expression for the former strategy has the form:
\begin{equation}
(\mu \nu \vert \rho \sigma)^x = \sum_P \left \{\frac{\partial L_{\mu \nu}^P}{\partial x} L_{\rho \sigma}^P + L_{\mu \nu}^P \frac{\partial L_{\rho \sigma}^P}{\partial x} \right \},
\end{equation}
and requires the differentiated CVs. While for magnetic perturbations it is no problem to handle these perturbed CVs~\cite{giao-mp2}, their handling is in case of geometrical perturbations cumbersome, as the CV elements depend (unlike the underlying two-electron integrals) on all perturbations. Strategies that avoid their construction~\cite{schnack2022efficient} are therefore the preferred way and again exploit the formal equivalence between the RI/DF and the CD approximations of the ERI matrix~\cite{pedersen2009density}. 

In RI/DF, a pre-optimized auxiliary basis set is introduced by a least-square fitting of the product densities $\vert \mu \nu )$:
\begin{equation}
    (\mu \nu \vert \rho \sigma ) \approx \sum_{QR}^{N_{aux}} (\mu \nu \vert Q) (Q \vert R)^{-1} (R \vert \rho \sigma),
\end{equation}
where $Q, R, ...$ are elements of the auxiliary basis, the $(\mu \nu \vert Q)$ are typically referred to as non-orthogonal vectors, and $(Q \vert R)$ is a metric matrix. 
One can perform the exact Cholesky factorization of the inverse of the metric to rewrite the non-orthogonal vectors as CVs:
\begin{equation}
     \label{eq:dfcd}
    (\mu \nu \vert \rho \sigma ) \approx \sum_{QR}^{N_{aux}} (\mu \nu \vert Q) (KK^T)_{QR}^{-1} (R \vert \rho \sigma) = \sum_{P} L_{\mu \nu}^P L_{\rho \sigma}^P,
\end{equation}
with:
\begin{equation}
    \label{eq:cv_2ndstep}
    L_{\rho \sigma}^P = \sum_R K_{PR}^{-1} (R \vert \rho \sigma).
\end{equation}
Taking the derivative of Equation~\ref{eq:dfcd}, we can reformulate the first derivative of the ERI tensor in terms of CVs, as first shown by Aquilante and co-workers~\cite{aquilante2008analytic}:
\begin{equation}
    \label{eq:eri_cddf_der}
    (\mu \nu \vert \rho \sigma)^x = \sum_{P} (\mu \nu \vert P)^x \tilde{L}_{\rho \sigma}^P + \sum_{P} \tilde{L}_{\mu \nu}^P(P \vert \rho \sigma)^x - \sum_{PQ} \tilde{L}_{\mu \nu}^P (P \vert Q)^x \tilde{L}_{\rho \sigma}^Q,
\end{equation}
where we have defined \textit{transformed} CVs $\tilde{L}_{\mu \nu}^P$:
\begin{equation}
    \tilde{L}_{\mu \nu}^P = \sum_Q K_{PQ}^{-T} L_{\mu \nu}^Q = \sum_Q K_{PQ}^{-T} \sum_R K_{QR}^{-1} (R \vert \mu \nu).
\end{equation}

\subsection{CCSD Lagrangian}

Due to the non-variational character of CC methods, a straightforward differentiation of the electronic energy would increase the computational complexity of the calculation. Writing the CC energy as $E = E(\mathbf{x},\mathbf{t})$, where $\mathbf{x}$ denotes the set of perturbation parameters and $\mathbf{t}$ the set of CC $t$-amplitudes, its first derivative takes the form:
\begin{equation}
    \label{eq:energy_der}
    \frac{d E}{d \mathbf{x}} = \frac{\partial E}{\partial \mathbf{x}} + \frac{\partial E}{\partial \mathbf{t}} \frac{\partial \mathbf{t}}{\partial \mathbf{x}} ,
\end{equation}
where the derivatives are evaluated at point $\mathbf{x} = \mathbf{0}$ and we ignore orbital relaxation for simplicity. Since the CC energy and wavefunction parameters are not determined variationally, the second term in the right-hand side of Equation~\ref{eq:energy_der} does not vanish. Therefore, it seems to become necessary to compute the derivatives of the $t$-amplitudes with respect to the perturbation parameters by solving the perturbed CC equations. For geometrical gradients with respect to nuclear coordinates this would be especially cumbersome, since an additional set of $3N_{atoms}$ linear equations with the same scaling as the unperturbed amplitude equations would have to be solved. 

The easiest way to eliminate the need to solve the perturbed amplitude equations is to define a CC Lagrangian~\cite{koch1990lagrangian} as follows:
\begin{equation}
    \mathcal{L}(\mathbf{x},\mathbf{t},\boldsymbol{\lambda},\mathbf{Z},\mathbf{I}) = \langle 0 \vert (1 + \hat{\Lambda})\mathcal{H} \vert 0 \rangle + 2\sum_{ai} Z_{ai} f_{ai} + \sum_{pq} I_{pq} \left( \sum_{\mu \nu} c_{\mu p} S_{\mu \nu} c_{\nu q} - \delta_{pq} \right),
\end{equation}
where $Z_{ai}$ and $I_{pq}$ are Lagrange multipliers used to enforce, respectively, Brillouin's condition ($f_{ai} =0$ with $f_{ai}$ as the corresponding Fock-matrix element) and orthonormality between MOs (given in terms of the AO overlap integrals $S_{\mu \nu}$), whereas $\hat{\Lambda}$ is a de-excitation operator defined as:
\begin{equation}
    \hat{\Lambda} = \hat{\Lambda}_1 + \hat{\Lambda}_2 + ... = \sum_{ai} \lambda_a^i \{\hat{i}^{\dagger}\hat{a}\} + \frac{1}{4} \sum_{abij} \lambda_{ab}^{ij} \{\hat{i}^{\dagger}\hat{a}\hat{j}^{\dagger}\hat{b}\} + ...
\end{equation}
which contains the so-called $\lambda$-amplitudes, i.e., another set of Lagrange multipliers used to enforce that the $t$-amplitudes satisfy the usual CC amplitude equations. As usual, indices $i$, $j$, ..., $m$, $n$, ... refer to occupied MOs, whereas indices $a$, $b$, ..., $e$, $f$, ... to virtual MOs. 
As the $\lambda$-amplitudes are Lagrange multipliers, there are as many of them as there are CC amplitude equations. It follows that the $\hat{\Lambda}$ operator expansion is naturally truncated at the same level as $\hat{T}$. Thus, for CCSD $\hat{\Lambda} = \hat{\Lambda}_1 + \hat{\Lambda}_2$.


\subsection{Stationarity with respect to cluster amplitudes: Lambda equations}

We require the CC Lagrangian to be stationary with respect to both $t$- and $\lambda$-amplitudes:
\begin{align}
    & \frac{\partial \mathcal{L}}{\partial \mathbf{t}} = 0, \\ 
    & \frac{\partial \mathcal{L}}{\partial \boldsymbol{\lambda}} = 0.
\end{align}
The latter equation is equivalent to the usual equations for the $t$-amplitudes and will not be discussed further here. However, the former stationarity condition yields a set of linear equations for the set $\boldsymbol{\lambda}$ of $\lambda$-amplitudes, usually referred to as Lambda equations.

Expressions for the singles and doubles Lambda equations as they are implemented are given in the following in a fully spin-adapted form for the closed-shell case (with an RHF reference wavefunction). Uppercase indices refer to $\alpha$ spin-orbitals, while lowercase indices to $\beta$ spin-orbitals. Our implementation is based on the equations reported by Gauss, Stanton, and Bartlett~\cite{gauss1991coupled}. 

The equations for the single $\lambda$-amplitudes read as:

\begin{equation}
\label{eq:lambda_singles}
    \begin{split}
        D_I^A\lambda_A^I =& \mathcal{F}_{IA} + \sum_e\lambda_E^I\mathcal{F}_{EA} - \sum_m\lambda_A^M\mathcal{F}_{IM} \\
        & +\sum_m\sum_e\lambda_e^m\left(2\widetilde{\mathcal{W}}_{EiMa} + \widetilde{\mathcal{W}}_{EimA}\right) \\
        & + 2\sum_P\mathcal{G}_PL^P_{ia} + \sum_P\sum_e\mathcal{G}^P_{EI}\left(L^P_{EA}-t^P_{EA}\right) \\
        & - \sum_{mn}\mathcal{G}_{mn}\left(2\mathcal{W}_{MiNa} - \mathcal{W}_{ImNa}\right) \\
        & - \sum_{mn}\sum_{e}\left(2\lambda_{Ae}^{Mn} - \lambda_{Ea}^{Mn}\right)\mathcal{W}_{IeMn} - \sum_m\mathcal{G}_{MI}\mathcal{F}_{MA} \\
        &+ \sum_P\sum_e v^P_{EI}L^P_{EA} + \sum_P\sum_m v^P_{MI}L^P_{MA} \\
        & - \sum_{mn}\sum_e\left(\Tilde{\Tilde{\mathcal{W}}}_{NeAm}\mathcal{V}_{ImNe}+\Tilde{\Tilde{\mathcal{W}}}_{NeaM}\mathcal{V}_{ImnE}\right),
    \end{split}
\end{equation}
and, accordingly, the equations for the double $\lambda$-amplitudes are:
\begin{equation}
   \label{eq:lambda_doubles}
    \begin{split}
        D_{Ij}^{Ab}\lambda_{Ab}^{Ij} = & \bra{Ij}\ket{Ab} \\
        & + P_-(ab)\sum_e\lambda_{Ae}^{Ij}\mathcal{F}_{eb} - P_-(ij)\sum_m\lambda_{Ab}^{Im}\mathcal{F}_{jm}\\
        & + \sum_{mn}\lambda_{Ab}^{Mn}\mathcal{W}_{IjMn} + \sum_{ef}\lambda_{Ef}^{Ij}\mathcal{W}_{EfAb} + \sum_{mn}\mathcal{V}_{MnIj}\bra{Mn}\ket{Ab}\\
        &+P_+(ia,jb)\lambda_A^I\mathcal{F}_{jb} + P_-(ab)\sum_m\lambda_A^M\mathcal{W}_{IjMb}\\
        &+\frac{1}{2}P_+(ia,jb)\sum_m\sum_e \left(2\lambda_{Im}^{Ae} - \lambda_{Im}^{Ea}\right)\left(2\Tilde{\mathcal{W}}_{MbEj}+\Tilde{\mathcal{W}}_{MbeJ}\right)\\
        &+ \frac{1}{2}P_+(ia,jb)\sum_m\sum_e \left(\frac{1}{2}\lambda_{Im}^{Ea}\Tilde{\mathcal{W}}_{EjMb}-\lambda_{Mi}^{Ae}\Tilde{\mathcal{W}}_{EjMb}\right)\\
        &+P_+(ia,jb)\sum_P\left(\mathcal{G}^P_{AI} - \mathcal{G}^P_{IA} + \lambda^P_{IA} - \Bar{\lambda}^P_{IA}\right)L^P_{jb},
    \end{split}
\end{equation}
All intermediates used in the solution of singles and doubles Lambda equations and appearing in Equations~\ref{eq:lambda_singles} and \ref{eq:lambda_doubles} are defined in appendix~\ref{appendix:lambda}.
Among these intermediates, those that do not depend on the $\lambda$-amplitudes are computed only once before starting the iterative solution of the Lambda equations and are stored in memory throughout it.

The Lambda equations are solved iteratively using the DIIS procedure~\cite{PULAY1980393,pulay1982diis} to accelerate convergence.

The rate-determining step in the iterative solution of the CCSD Lambda equations is the contraction within the particle-particle ladder (PPL) contribution:
\begin{equation}
    Z_{Ab}^{Ij} = \sum_{ef} \lambda_{Ef}^{Ij} \mathcal{W}_{EfAb},
\end{equation}
which has a formal $\mathcal{O}(O^2V^4)$ scaling, with $O$ and $V$ being number of occupied and virtual MOs, respectively. We follow here the same strategy that we have employed for the PPL term appearing in the double $t$-amplitude equations~\cite{Nottoli2023}. In order to reduce its computational cost from its formal $O^2V^4$ to $\tfrac{1}{4}O^2V^4$, we applied the well-known symmetric-antisymmetric algorithm~\cite{saebo1987fourth}. The product of the contraction $Z_{Ab}^{Ij}$ is written as the sum of a symmetric ($S_{Ab}^{Ij}$) and an antisymmetric ($A_{Ab}^{Ij}$) part:
\begin{equation}
    Z_{Ab}^{Ij} = S_{Ab}^{Ij} + A_{Ab}^{Ij},
\end{equation}
where
\begin{align}
    &S_{Ab}^{Ij} = \sum_{ef} \prescript{+}{}{\lambda_{Ef}^{Ij}} \prescript{+}{}{\mathcal{W}_{EfAb}},\\
    &A_{Ab}^{Ij} = \sum_{ef} \prescript{-}{}{\lambda_{Ef}^{Ij}} \prescript{-}{}{\mathcal{W}_{EfAb}},\\
    &\prescript{\pm}{}{\lambda_{Ef}^{Ij}} = \frac{1}{2} \left(\lambda_{Ef}^{Ij} \pm \lambda_{Fe}^{Ij}\right),\\
    &\prescript{\pm}{}{\mathcal{W}_{EfAb}} = \frac{1}{2} \left(\mathcal{W}_{EfAb} \pm \mathcal{W}_{FeAb}\right).
\end{align}
By exploiting the permutational symmetries of $\prescript{\pm}{}{\lambda_{Ef}^{Ij}}$ and $\prescript{\pm}{}{\mathcal{W}_{EfAb}}$, we are free to store in memory only the elements of the two tensor pairs whose indices satisfy the constraints $A \geq b$, $I \geq j$ and $E \geq f$ and, likewise, the sum of the PPL contraction is allowed to run over the $E \geq f$ indices. Since this has to be performed for both the symmetric and antisymmetric contributions, a $\tfrac{1}{4}$ pre-factor is gained. In order to avoid storing $V^4$ and $V^3O$ scaling arrays, such as the full $\prescript{\pm}{}{\mathcal{W}_{EfAb}}$ intermediates, for a memory-efficient implementation, we keep the $a$ index fixed by means of an external loop over virtual indices. Moreover, this $a$ loop is parallelized and the operations nested within are distributed over shared-memory threads through the OpenMP directive, so that we store at most $V^3N_{threads}$ temporary quantities. 

\subsection{CCSD density matrices}

The final CC gradient expression is usually written in terms of density matrices~\cite{Rice85,Bartlett86} in order to separate its perturbation-dependent and perturbation-independent constituents. The perturbation dependence is then entirely due to the integral derivatives, while
the perturbation-independent part is used to define the CC density matrices, which, in a second quantization formalism, take the form:
\begin{align}
   &D_{pq} = \langle 0 \vert (1 + \hat{\Lambda}) e^{-\hat{T}} \{\hat{p}^{\dagger} \hat{q} \} e^{\hat{T}} \vert 0 \rangle,\\
   &\Gamma_{pqrs} = \langle 0 \vert (1 + \hat{\Lambda}) e^{-\hat{T}} \{\hat{p}^{\dagger} \hat{q}^{\dagger} \hat{s} \hat{r} \}  e^{\hat{T}} \vert 0 \rangle.
\end{align}
In particular, the CCSD one-body and two-body density matrices are constructed by means of converged $t$- and $\lambda$-amplitudes. 
Expressions for the spin-adapted blocks of the CCSD one- and two-body density matrices,
along with the intermediates used to compute them, are reported in appendix~\ref{appendix:d1cc} and appendix~\ref{appendix:d2cc}.
It should, however, be noted that, when taking into account orbital-relaxation effects, the particle-hole block of the one-body density matrix does not need to be computed.

The $\mathcal{G}$ and $\mathcal{V}$ intermediates (defined in appendices~\ref{appendix:lambda} and \ref{appendix:d2cc}) are calculated prior to the construction of the density matrices and subsequently stored in memory. Nonetheless, quantities with a $V^4$ or $V^3O$ memory scaling ($\Gamma_{AbCd}$, $\Gamma_{AbCi}$ and $\mathcal{V}_{AbEf}$) are never explicitly constructed and stored, but are contracted on-the-fly with the appropriate tensors when needed. A noteworthy observation can be made concerning the non-Hermitian nature of the CC Lagrangian: in order to guarantee real-valued results, only the Hermitian components of the CC density matrices need to be evaluated~\cite{Shavitt_Bartlett_2009}.  Furthermore, for calculations with an RHF reference, only the spin-adapted form of the two-body density matrix is computed and stored:
\begin{equation}
    \widetilde{\Gamma}_{pqrs} = 2\Gamma_{PqRs} - \Gamma_{PqSr}.
\end{equation}

\subsection{Stationarity with respect to orbital rotations: $Z$-vector equations and expressions for the $I_{pq}$ intermediates}

Orbital relaxation following a perturbation can be parameterized by rewriting the perturbed molecular orbital coefficients as a linear combination of the unperturbed coefficients~\cite{pople79}:
\begin{equation}
    C_{\mu p}^x = \sum_q C_{\mu q} U_{qp}^x,
\end{equation}
where the $U_{qp}^x$ coefficients are solutions of the coupled-perturbed Hartree-Fock (CPHF) equations~\cite{Gerrat68,pople79}. By enforcing stationarity of the CC Lagrangian with respect to the various blocks of the $U_{qp}^x$ matrix:
\begin{align}
    &\frac{\partial \mathcal{L}}{\partial U_{ij}^x} = 0,\\
    &\frac{\partial \mathcal{L}}{\partial U_{ab}^x} = 0,\\
    &\frac{\partial \mathcal{L}}{\partial U_{ia}^x} = 0,\\
    &\frac{\partial \mathcal{L}}{\partial U_{ai}^x} = 0,
\end{align}
expressions for the Lagrange multipliers $I_{ij}$, $I_{ab}$, $I_{ai} + I_{ia}$ and $Z_{ai}$ are obtained. The fourth stationarity condition, in particular, leads to the $Z$-vector equations \cite{Handy84}:
\begin{equation}
  \label{eq:z_vector}
  \sum_m \sum_e Z_{em} \left[ (\varepsilon_a - \varepsilon_i)\delta_{ea} \delta_{mi} + 2\langle Ei \vert Ma \rangle - \langle Ei \vert Am \rangle + 2\langle Ea \vert Mi \rangle - \langle Ea \vert Im \rangle \right] = 2\left(I^{\prime}_{IA} - I^{\prime}_{AI}\right),    
\end{equation}
where:
\begin{equation}
    I^{\prime}_{PQ} = -\varepsilon_P D_{PQ} - 4 \sum_{rst} \langle Pr \vert St \rangle\widetilde{\Gamma}_{qrst} -\sum_{rs} (2 \langle Pr \vert Qs \rangle - \langle Pr \vert Sq \rangle) D_{rs} \delta_{q,occ.}
\end{equation}
It should be noted that the third term in $I^{\prime}_{PQ}$ needs to be computed only when $q$ is a hole index.
By solving the linear system in Equation~\ref{eq:z_vector}, one can define a relaxed one-body density matrix $D'_{pq}$:
\begin{align}
    &D'_{ij} = D_{ij},\\
    &D'_{ab} = D_{ab},\\
    &D'_{ia} = Z_{ai},\\
    &D'_{ai} = Z_{ai}.
\end{align}

As for the spin-adapted $I_{pq}$ multipliers, they can be expressed as:
\begin{align}
  &I_{AB} = I_{AB}^{\prime},\\
  &I_{AI} = I_{IA}^{\prime} - \varepsilon_I Z_{AI},\\
  &I_{IA} = I_{IA}^{\prime} - \varepsilon_I Z_{AI},\\
  &I_{IJ} = I_{IJ}^{\prime} - \sum_m \sum_{e} \left[(2 \langle Jm \vert Ie \rangle - \langle Jm \vert Ei \rangle) + (2 \langle Im \vert Je \rangle - \langle Im \vert Ej \rangle) \right] Z_{em}.
\end{align}


The symmetric-antisymmetric algorithm has been applied here to all $\mathcal{O}(O^2V^4)$ scaling contractions, thus reducing the number of associated floating-point operations by a factor of four. Furthermore, a modified version of the same algorithm, which we will refer to as partial symmetric-antisymmetric algorithm, has been implemented for all $\mathcal{O}(O^3V^3)$ scaling contractions, reducing their computational cost by a factor of two. For instance, by looking at the following contraction:
\begin{equation}
    M_{bi}^{dc} = \sum_{mn} \lambda_{Bi}^{Mn} \tilde{\tau}_{mn}^{dc},
\end{equation}
in the "partial" algorithm we are limited to the construction of tensors of the type $\prescript{\pm}{}{\lambda_{Bi}^{M \geq n}}$, due to the fact that $\lambda_{Bi}^{Mn}$ is not a square matrix with respect to the $Bi$ pair of indices.

\subsection{CCSD analytic first derivatives}

Due to the stationarity of the CC Lagrangian with respect to the CCSD amplitudes and Lagrange multipliers, the total first derivative of the energy with respect to a generic perturbation $x$ can be written as follows:
\begin{equation}
    \label{eq:gradcc_mo}
    \frac{dE}{dx} = \frac{\partial \mathcal{L}}{\partial x} = 2 \sum_{pq} D'_{pq} f_{pq}^{(x)} + \sum_{pqrs} \widetilde{\Gamma}_{pqrs} \langle Pq \vert Rs \rangle^x + 2 \sum_{pq} I_{pq} S_{pq}^x,
\end{equation}
where the derivatives of the Fock matrix, the overlap matrix and the two-electron integrals are evaluated in the AO basis and then transformed into the MO basis:
\begin{equation}
  f_{pq}^{(x)} = \sum_{\mu \nu} C_{\mu p} f_{\mu \nu}^{(x)} C_{\nu q}.
\end{equation}
The response of the MO coefficients is here not included, as it has been accounted for in the treatment of orbital relaxation via the $Z$-vector equations.

The contractions between the CC density matrices/the $I_{pq}$ intermediates and the differentiated integrals are actually performed in the AO basis, as to avoid the storage of the memory-intensive two-electron integral derivatives in the MO basis and their repeated four-index transformation from the AO to the MO basis. Equation~\ref{eq:gradcc_mo} thus takes the form:
\begin{equation}
    \label{eq:gradcc_ao}
    \frac{dE}{dx} = 2 \sum_{\mu \nu} D'_{\mu \nu} f_{\mu \nu}^{(x)} + \sum_{\mu \nu \rho \sigma} \widetilde{\Gamma}_{\mu \rho \nu \sigma} (\mu \nu \vert \rho \sigma)^x + 2 \sum_{\mu \nu} I_{\mu \nu} S_{\mu \nu}^x.
\end{equation}
The derivatives of the Fock and overlap matrices are directly built and stored in the AO basis, thus the one-body density matrix and the $I_{pq}$ multiplier are transformed into the AO basis prior to the contraction.

The full transformation of the two-body density matrix into the AO basis, on the other hand, would lead to the storage of a conspicuous 
$N^4$ quantity. To avoid this issue, we exploit the formal equivalence between RI/DF and CD, as already mentioned in subsection~\ref{subsection:cd}.
In this way, it becomes possible to rewrite the second term in Equation~$\ref{eq:gradcc_ao}$ by substituting in the expression given in Equation~\ref{eq:eri_cddf_der} as follows:
\begin{equation}
    \label{eq:2el_grad_cont}
    \sum_{\mu \nu \rho \sigma} \widetilde{\Gamma}_{\mu \rho \nu \sigma} (\mu \nu \vert \rho \sigma)^x = \sum_P (\mu \nu \vert P)^x J_{\mu \nu}^P + \sum_P \Bar{J}_{\rho \sigma}^P (P \vert \rho \sigma)^x - \sum_{PQ} W_{PQ} (P \vert Q)^x,
\end{equation}
where the following intermediates have been defined and exploited:
\begin{align}
    &\tilde{L}_{qs}^P = \sum_{\rho \sigma} C_{\rho q} C_{\sigma s} \tilde{L}_{\rho \sigma}^P,\\
    &\Gamma_{pr}^P = \sum_{qs} \widetilde{\Gamma}_{pqrs}\tilde{L}_{qs}^P, \label{eq:d2xtl1}\\
    &\bar{\Gamma}_{qs}^P = \sum_{pr} \widetilde{\Gamma}_{pqrs}\tilde{L}_{pr}^P, \label{eq:d2xtl2}\\
    &J_{\mu \nu}^P = \sum_{pr} C_{\mu p} C_{\nu r} \Gamma_{pr}^P,\\
    &\bar{J}_{\rho \sigma}^P = \sum_{qs} C_{\rho q} C_{\sigma s} \bar{\Gamma}_{qs}^P,\\
    &W_{PQ} = \sum_{\mu \nu} \tilde{L}_{\mu \nu}^P J_{\mu \nu}^Q.
\end{align}
This implementation makes use of two- and three-index density intermediates (with the most memory-expensive one scaling as $\mathcal{O}(N^2N_{ch})$), which are contracted on-the-fly with differentiated two-electron integrals. Integral derivatives are generated by the {\sc MINT} package~\cite{gauss2015mint} within {\sc CFOUR}~\cite{cfour} in batches: given a generic integral $(\mu \nu \vert \rho \sigma)^x$, if either of the two product densities $\vert \mu \nu)$ or $\vert \rho \sigma )$ corresponds to a Cholesky index, it follows that the currently considered integral is the derivative of a non-orthogonal vector and is contracted with the sum $J_{\mu \nu}^P + \Bar{J}_{\mu \nu}^P$ (or $J_{\rho \sigma}^P + \Bar{J}_{\rho \sigma}^P$); if both correspond to Cholesky indices, the integral is the derivative of the metric and is therefore contracted with $W_{PQ}$. It should be noted that each block of the CCSD two-body density matrix is considered separately and all contributions are sequentially added to the $J_{\mu \nu}^P$ and $\Bar{J}_{\mu \nu}^P$ intermediates. Furthermore, when considering symmetric blocks of the two-body density matrix, $J_{\mu \nu}^P = \Bar{J}_{\mu \nu}^P$, thus only one of the two is actually computed. As usual, in Equations~\ref{eq:d2xtl1} and \ref{eq:d2xtl2} the contractions involving the $vvvo$ and $vvvv$ blocks of $\widetilde{\Gamma}_{pqrs}$ are performed on-the-fly.

\section{Computational details}

\label{section:comput}

\subsection{Exploitation of Abelian point-group symmetry}

A main aspect of our implementation of CCSD analytic gradients is that it fully exploits Abelian point-group symmetry by means of the direct product decomposition (DPD) scheme~\cite{stanton1991direct}. Both theoretical arguments and benchmark calculations have shown that, when explicitly including Abelian symmetry in CC calculations, timings can be reduced by a factor of $h^2$, where $h$ is the order of the point group of the considered molecule. 

Molecular orbitals naturally transform as the irreducible representations of the molecular point group, whereas atomic orbitals have to be linearly combined into a set of symmetry-adapted basis functions (SALCs). Therefore, all tensors appearing in our implementation satisfy the usual symmetry selection rules. In particular, it suffices that the direct product of the irreducible representations of all indices for each quantity equals the totally symmetric representation (for Abelian point groups, given all irreducible representations are one-dimensional), so that only non-vanishing elements are evaluated and stored. Contractions that involve two four-index matrices that share a pair of indices, e.g. the PPL term, are thus divided into $h$ smaller operations.

Moreover, when evaluating geometrical gradients, {\sc CFOUR} computes the derivatives of one-electron and two-electron integrals with respect to symmetry-adapted nuclear displacements, thereby using double cosets to construct the symmetry-adapted integrals as described by Davidson~\cite{Davidson75} and Taylor~\cite{Taylor86} in such a way that the overall differentiated integral matrix is itself totally symmetric. The symmetry-adapted gradient is later back-transformed into Cartesian coordinates.

\subsection{Symmetry-adapted two-step CD algorithm}

Our code exploits a two-step algorithm for the CD of ERIs and their derivatives, which was originally proposed by Aquilante \textit{et al.}~\cite{Aquilante2011}, and later improved by Folkestad \textit{et al.}~\cite{folkestad2019twostep}. In particular, our implementation is inspired by the one described by Zhang \textit{et al.}~\cite{zhang2021toward}. This two-step framework leads to a reduction in the number of FLOPs and RAM requirements with respect to the conventional algorithm, by dividing the decomposition procedure into two subtasks: in the first step, the Cholesky basis is determined, without computing the full CVs, since only diagonal elements of the ERI matrix need to be evaluated and elements that give a negligible contribution are discarded; in the second step, the CVs are obtained directly by means of dense linear algebra operations that can be performed using highly optimized BLAS and LAPACK routines, in an analogous fashion as RI/DF. Moreover, in accordance with the rest of the CD-CCSD code, our implementation of the two-step algorithm fully exploits Abelian point-group symmetry. 

The first step, where Cholesky pivoting elements are determined, is summarized in Algorithm~\ref{alg:firststep_cd}. The algorithm depends on three parameters only: $\tau$, which is a user-provided threshold controlling the approximation error, $\sigma$, generally known as the "span factor", which defines the lists of qualified Cholesky pivots, and $Q_{max}$, which limits the maximum size of batches of qualified elements. $\sigma$ and $Q_{max}$ are set equal to $10^{-2}$ and 1000 respectively, as suggested by Folkestad \textit{et al.}~\cite{folkestad2019twostep}. In the second step, CVs are computed through Equation~\ref{eq:cv_2ndstep}, where efficient BLAS and LAPACK routines can be used for the factorization of the Cholesky basis metric and the evaluation of the contraction itself.

\begin{algorithm}
\caption{Symmetry-adapted Cholesky basis determination} \label{alg:firststep_cd}
\begin{algorithmic}[1]
  \State Define a list $\mathcal{P}$ to store significant basis pairs
  \State Define a list $\mathcal{D}$ to store significant diagonal elements
  \State Define lists $\mathcal{L}(\Gamma)$ to store Cholesky vectors based on the \textit{irrep} $\Gamma$ of the pivots
  \For{$R = AB, A \geq B$ shell pairs}
      \State Compute the diagonal shell quartet $(R \vert R)$
      \For{$r \in R, r = \mu \nu, \mu \geq \nu$}
          \If{$\text{Diag}(r) = (r \vert r) \geq \tau$}
              \State Add $r$ to list $\mathcal{P}$ $(\mathcal{P} \gets r)$
              \State Add $\text{Diag}(r)$ to list $\mathcal{D}$ $(\mathcal{D} \gets \text{Diag}(r))$
          \EndIf
      \EndFor
  \EndFor
  \While{$\mathcal{P} \neq \emptyset$}
    \State Take the $Q_{max}$ largest elements $r \in \mathcal{P}$ such that Diag$(r) \geq \sigma D_{max}$
    \State Add the selected elements to lists $\mathcal{Q}(\Gamma_r)$ based on their \textit{irrep} $\Gamma_r$
    \For{$r \in \mathcal{Q}$}
      \For{$s \in \mathcal{P}$}
        \State Compute the shell quartet $(R \vert S)$, with $r \in R$, $s \in S$, if not already allocated
        \For{$t \in R, u \in S, t, u \in \mathcal{P}$}
          \State Add the integrals $(t \vert u)$ and $(u \vert t)$ to lists $\mathcal{I}(\Gamma_t)$ based on the \textit{irrep} $\Gamma_t = \Gamma_u$
          \State $\tilde{M}_{tu} (\Gamma_t) = (t \vert u) - \sum_{L^p \in \mathcal{L}(\Gamma_p)} L_t^p L_u^p$ 
        \EndFor
      \EndFor 
    \EndFor
    \While{Diag$(q) \geq \tau_{mic}$, with Diag$(q) = \max_{r \in \mathcal{Q}} \text{Diag}(r)$}
        \State Take $q$ as pivoting element
        \For{$r \in \mathcal{P}$}
          \If{$\Gamma_r = \Gamma_q$}
            \State $L_r^q = \tfrac{\tilde{M}_{rq}(\Gamma_q) - \sum_{L^p \in \mathcal{L}_{mic}(\Gamma_q)}L_r^p L_q^p}{\sqrt{\text{Diag}(q)}}$ 
            \State Diag$(r)$ = Diag$(r) - \left( L_r^q \right)^2$
          \EndIf
        \EndFor
        \State Add $L^q$ to list $\mathcal{L}_{mic}(\Gamma_q)$
    \EndWhile
    \State $\mathcal{L}(\Gamma) = \mathcal{L}(\Gamma) \cup \mathcal{L}_{mic}(\Gamma)$ for all \textit{irreps}
    \For{$r \in \mathcal{P}$}
      \If{Diag$(r) < \tau$}
        \State $\mathcal{P} = \mathcal{P} \setminus \{ r \}$
        \State $\mathcal{D} = \mathcal{D} \setminus \{ \text{Diag}(r) \}$
        \State $\mathcal{I}(\Gamma_r) = \mathcal{I}(\Gamma_r) \setminus \{ (r \vert s), (s \vert r)\}$ for all $s \in \mathcal{P}$ such that $\Gamma_r = \Gamma_s$
        \State Remove the element $r$ from Cholesky vectors $L \in \mathcal{L}(\Gamma_r)$
      \EndIf
    \EndFor
  \EndWhile
\end{algorithmic}
\end{algorithm}



\section{Test calculations and discussion}

\label{section:test}

To test and demonstrate the efficiency of our implementation, we have performed geometry optimizations of two medium-large symmetric systems, coronene (C$_{24}$H$_{12}$) and hexabenzocoronene (C$_{42}$H$_{18}$), using geometrical gradients computed analytically at the CD-CCSD level of theory. The calculations were carried out with a development version of {\sc CFOUR}~\cite{code_cfour}. We used the default optimizer implemented within {\sc CFOUR}, which exploits a quasi-Newton scheme with a Broyden-Fletcher-Goldfarb-Shanno (BFGS) update~\cite{nocedal2006,fischer1992_qn}, where the identity matrix is chosen as initial guess for the Hessian. The optimization is considered converged as soon as the root mean square (RMS) of the gradient is lower than $10^{-5}$ Hartree/Bohr. We used the following thresholds for the aforementioned calculations: $10^{-7}$ for the convergence of the iterative SCF procedure, $10^{-8}$ for the convergence of the solution of the CC amplitude and Lambda equations, $10^{-12}$ for the convergence of the $Z$-vector equations and $\tau = 10^{-4}$ as tolerance for the CD of the ERI matrix. In particular, this last choice is justified by the fact that a less tight tolerance value for the CD threshold, namely $10^{-3}$, may cause slow convergence of the geometry optimization 
at the chosen convergence criteria of the optimizer, since the Cholesky basis varies along the PES, thus inducing discontinuities of the order of magnitude of $\tau$, as recognized by Schnack-Petersen \textit{et al.}~\cite{schnack2022efficient} and Aquilante \textit{et al.}~\cite{aquilante2008analytic}. Furthermore, as noted by Feng \textit{et al.}~\cite{feng2019implementation}, a Cholesky threshold of $10^{-4}$ leads to errors in the gradient and in the optimized geometries that are lower than the intrinsic accuracy of the CCSD method. Another aspect worth mentioning is that all electrons (both core and valence) are explicitly correlated in the reported calculations. 

\subsection{Analysis of timings}

\begin{figure}
    \centering
    \includegraphics[width=0.5\linewidth]{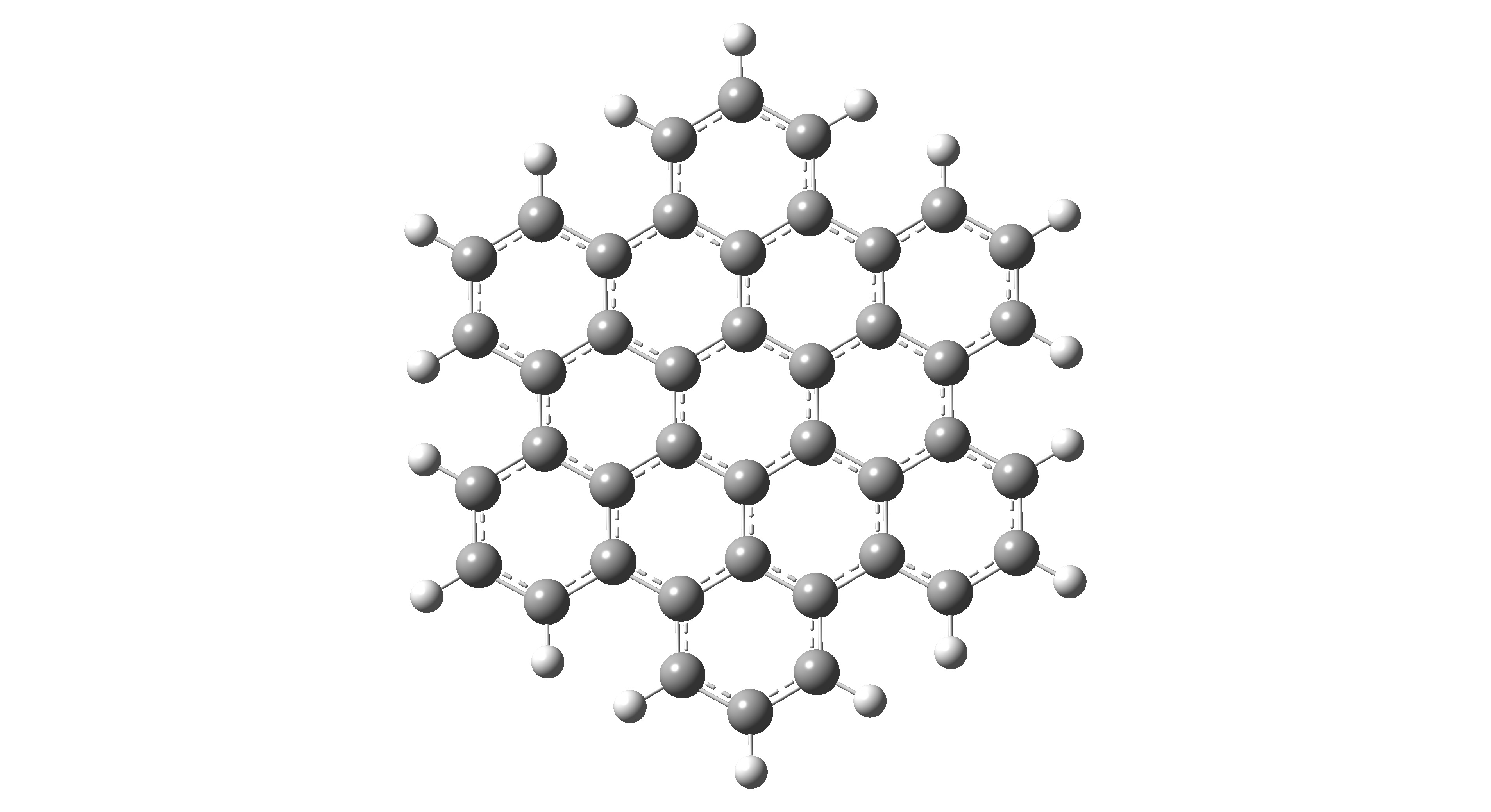}
    \caption{Structure of the hexabenzocoronene molecule, as used in its geometry optimization.}
    \label{fig:hbc_geometry}
\end{figure}

\begin{figure}[h]
\centering
\includegraphics[scale=0.6,angle=0]{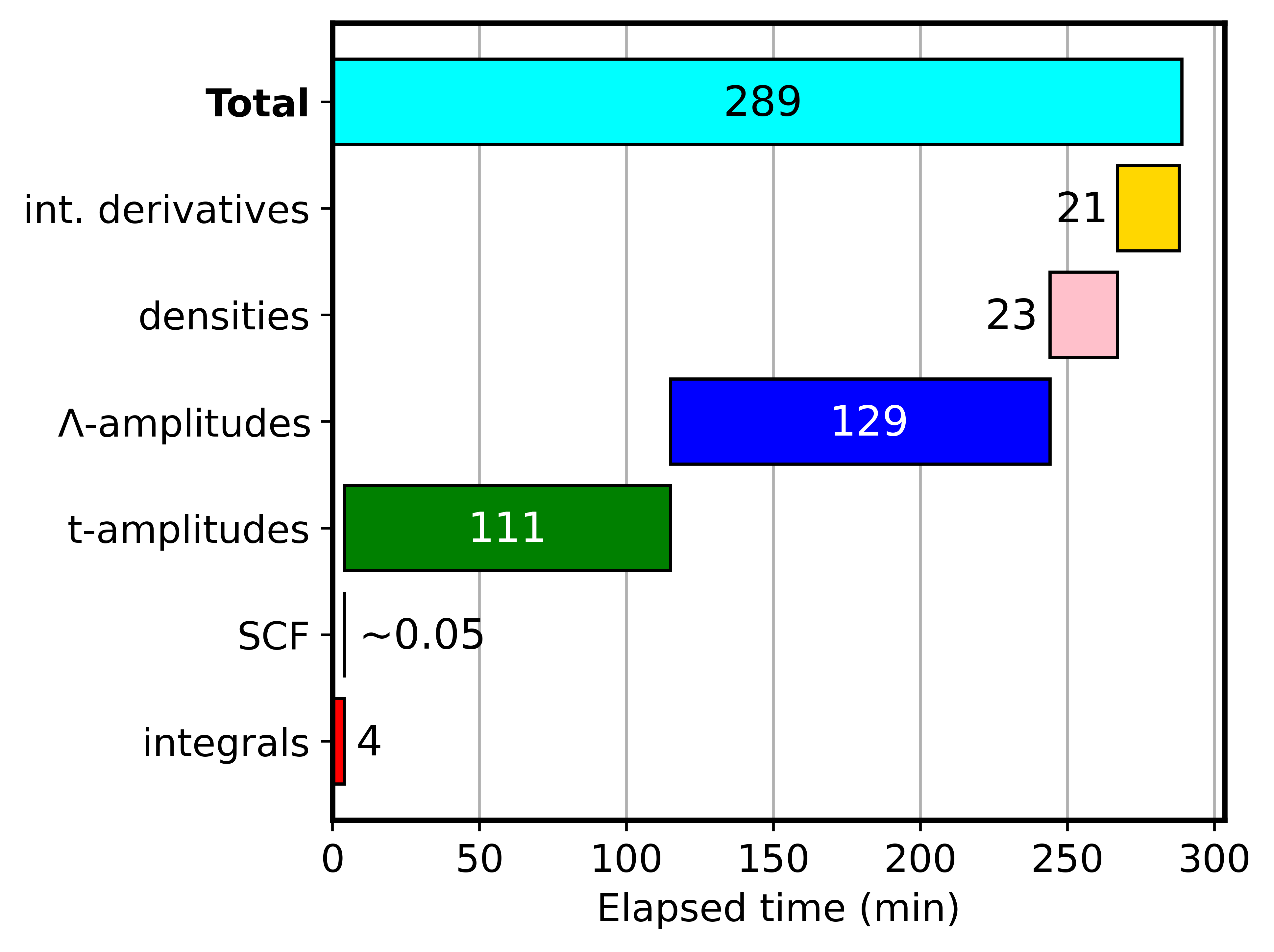}
\caption{Gantt chart for the timings (in minutes) for each step necessary for the calculation of CD-CCSD gradients of hexabenzocoronene using the cc-pVDZ basis.}
\label{fig:gantt_hbc}
\end{figure}

We have performed the geometry optimization of hexabenzocoronene (shown in Figure~\ref{fig:hbc_geometry}) on a node with an AMD EPYC 7282 16 Core processor, equipped with 512 GB of RAM and requesting 32 OpenMP threads. The calculation was carried out using Dunning's cc-pVDZ basis set~\cite{Dunning1989}, which consists of 678 basis functions, 135 occupied MOs, and 543 virtual MOs, and the enforced point group in the computation was $D_{2h}$ as the largest Abelian subgroup of $D_{6h}$, the full point group of the initial geometry. The two-step CD algorithm yielded 3710 CVs on average during the optimization, almost evenly distributed among the 8 irreducible representations. The optimization procedure converged to the equilibrium geometry in 24 steps, taking about 4 days and 15 hours in total. The timings associated with the individual tasks in the computation of gradients in the first optimization step are shown in the Gantt chart in Figure~\ref{fig:gantt_hbc}.

\begin{figure}
    \centering
    \includegraphics[width=0.5\linewidth]{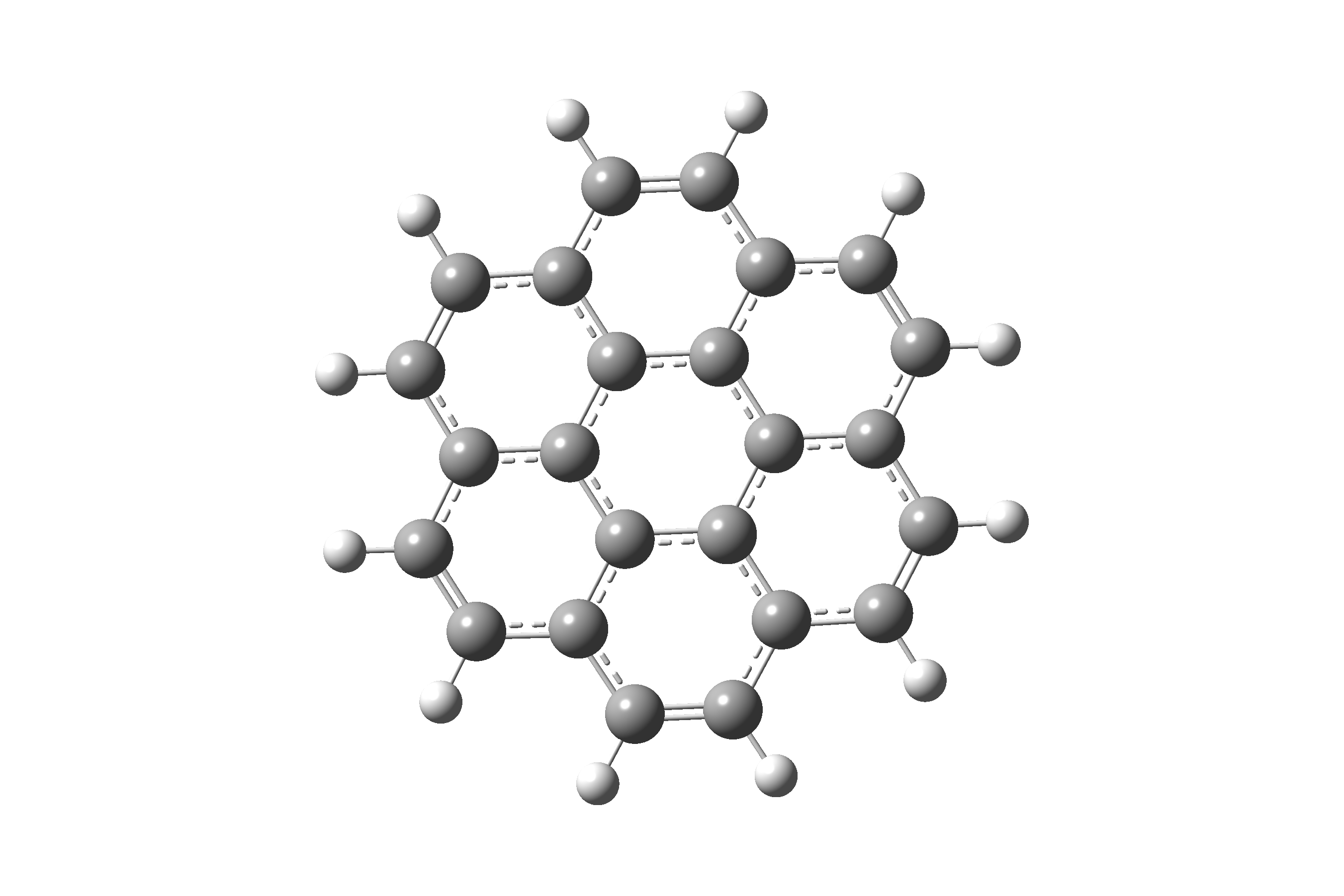}
    \caption{Structure of the coronene molecule, as used in its geometry optimization.}
    \label{fig:coronene_geometry}
\end{figure}

\begin{figure}[h]
\centering
\includegraphics[scale=0.6,angle=0]{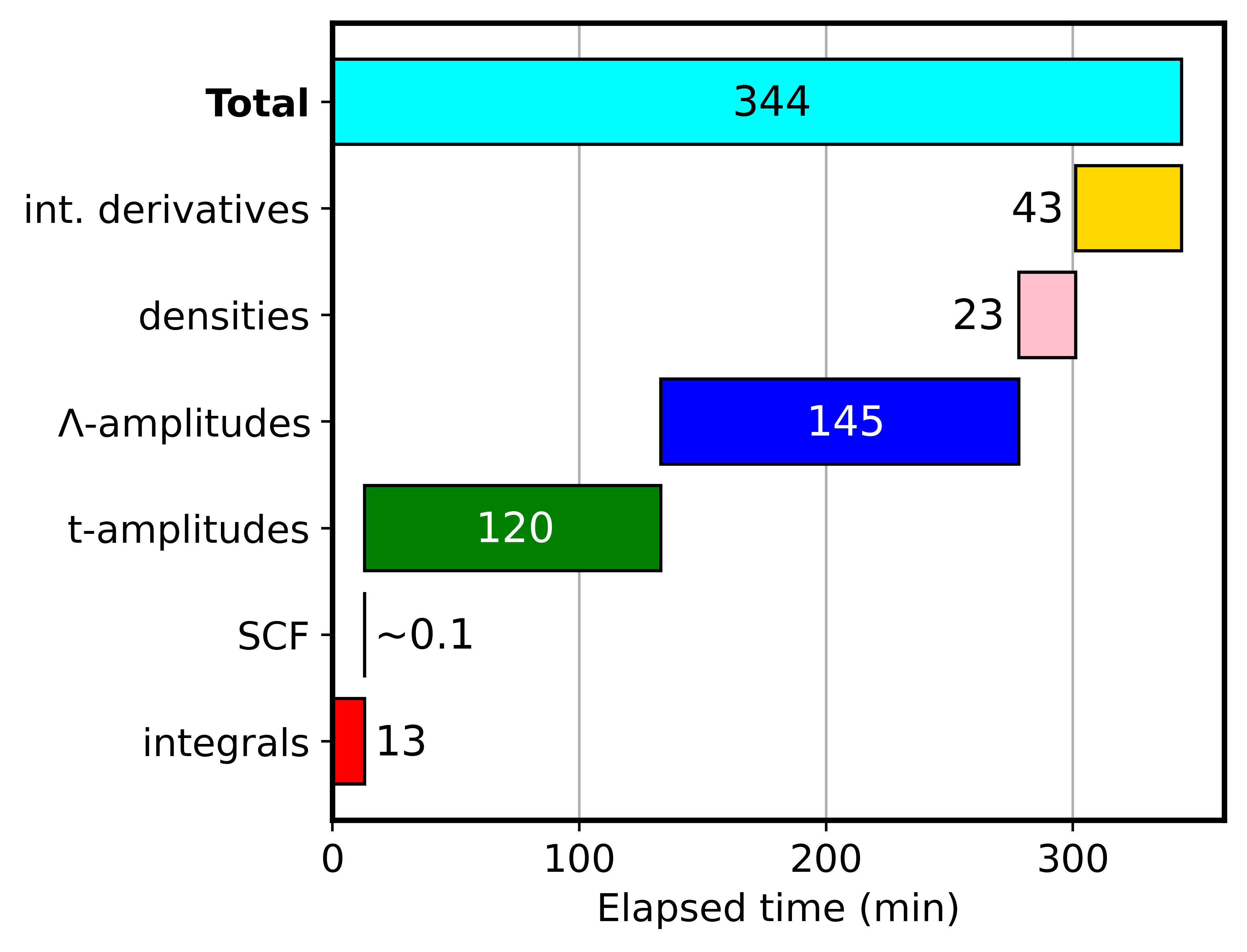}
\caption{Gantt chart for the timings (in minutes) for each step necessary for the calculation of CD-CCSD gradients of cc-pVTZ coronene.}
\label{fig:gantt_coronene}
\end{figure}

Furthermore, we have performed a geometry optimization for the coronene molecule (shown in Figure~\ref{fig:coronene_geometry}) on an Intel Xeon Gold 6140M node, equipped with 1280 GB of RAM and requesting 32 OpenMP threads. The calculation was carried out in Dunning's cc-pVTZ basis set~\cite{Dunning1989}, involving 888 basis functions, 78 occupied MOs, and 810 virtual MOs, and the enforced point group in the computation was $D_{2h}$, for the same reason as in the case of hexabenzocoronene. The two-step CD algorithm produced 4420 CVs on average during the optimization. The optimization procedure converged to the equilibrium geometry in 8 steps, taking about 2 days in total. The timings associated with the individual tasks in the computation of gradients in the first optimization step are shown in the Gantt chart in Figure~\ref{fig:gantt_coronene}.

Interestingly, the percentage of total elapsed time dedicated to the computation of integrals and integral derivatives increases by expanding the basis-set size. This is consistent with the fact that the computational gain due to Abelian point-group symmetry is comparable to the order of the point-group for the construction of integrals against a reduction by a factor of the square of the order of the point-group for CC calculations.

\subsection{Parallelization speedup analysis}

\begin{figure}
    \centering
    \includegraphics[scale=0.6]{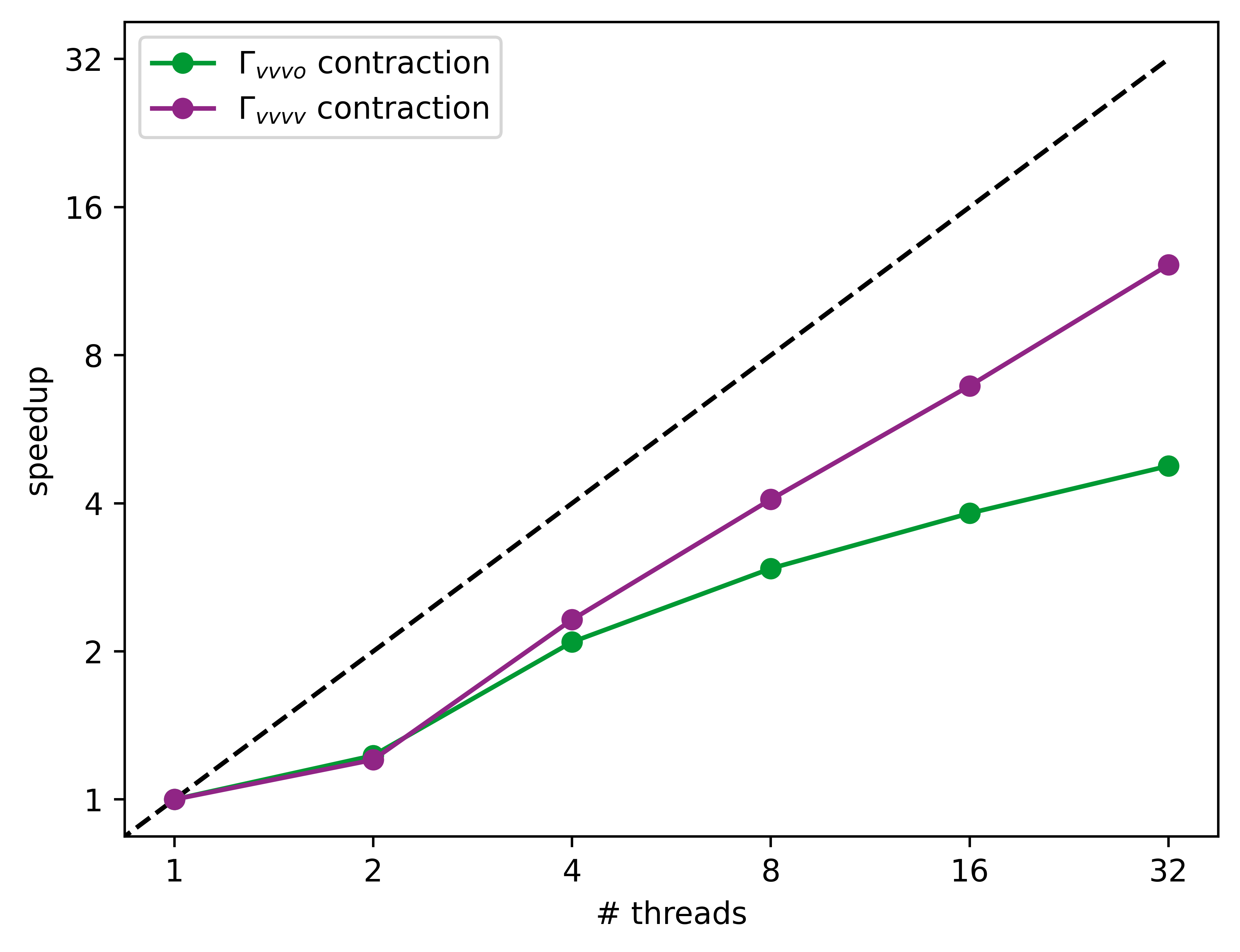}
    \caption{Speedup graph for CD-CCSD/cc-pVDZ gradient calculations on coronene. The green curve refers to the contraction involving the \textit{vvvo} block of the two-body density matrix in the evaluation of the molecular gradient, while the purple curve refers to the analogous contraction involving the \textit{vvvv} block of the two-body density matrix. The ratio between the execution time of the serial code and the execution time of the parallel one is plotted against the number of used OpenMP threads, in the log$_2$ scale.}
    \label{fig:speedup_coronene}
\end{figure}

In order to test the efficiency of the parallelization of the CD-CCSD gradient evaluation, we have performed several calculations on the coronene molecule using the cc-pVDZ basis on an AMD EPYC 7282 16 Core processor node, each employing a different number of shared-memory OpenMP threads (in particular, 1, 2, 4, 8, 16, and 32). The ratio between the wall time for the execution of the serial code over the wall time for the execution of the parallel code is plotted in Figure~\ref{fig:speedup_coronene}. The present analysis has been carried out on the contractions between the Cholesky-decomposed two-electron integral derivatives and the \textit{vvvo} (reported in the green curve) and the \textit{vvvv} (reported in the purple curve) blocks of the CCSD two-body density matrix, implemented as shown in Equation~\ref{eq:2el_grad_cont}.

The most expensive terms in the contraction involving the \textit{vvvo} block of $\Gamma$ are the following:
\begin{align}
    &\Gamma_{ac}^P = - \frac{1}{2} \sum_{bi} \sum_e \sum_{mn} \tilde{\tau}_{mn}^{ab} \lambda_{Ce}^{Mn} t_i^e \tilde{L}_{bi}^P, \\
    &\Bar{\Gamma}_{bi}^P = - \frac{1}{2} \sum_{ac} \sum_e \sum_{mn} \tilde{\tau}_{mn}^{ab} \lambda_{Ce}^{Mn} t_i^e \tilde{L}_{ac}^P,
\end{align}
originating from Equations~\ref{eq:d2xtl1} and \ref{eq:d2xtl2} and the third term in Equation~\ref{eq:tpdm_vvvo}. These have a formal $\mathcal{O}(O^3V^3)$ scaling, reduced by a factor of two by virtue of the partial symmetric-antisymmetric algorithm. To avoid storing $OV^3$ intermediates, these terms are evaluated within a parallelized external loop over the $c$ virtual index. The speedup plot related to such contractions does not show an ideal behavior, seemingly approaching a plateau as soon as with 16 OpenMP threads, likely due to the complexity of the implementation of the $\Gamma_{abci}$ contraction, which features relevant portions of serial code.

As for the contraction involving the \textit{vvvv} block of $\Gamma$, the most expensive term takes the following form:
\begin{equation}
    \mathcal{V}_{AbCd} = \sum_{mn} \tilde{\tau}_{mn}^{cd} \lambda_{Ab}^{Mn},
\end{equation}
which scales as $\mathcal{O}(O^2V^4)$. The contraction is performed within a parallelized loop over the $a$ virtual index, which is thus kept fixed when evaluating the matrix-matrix product, and the resulting intermediate is immediately contracted with a transformed CV in the MO basis. It should be observed that the full symmetric-antisymmetric algorithm cannot be applied in this case. The speedup plot shows the expected linear scaling when increasing from 2 up to 32 threads, even though the speedup values themselves deviate from ideality starting from 2 threads, possibly due to the few serial sections in the code. Furthermore, there is no sign of approaching a plateau even when requesting 32 OpenMP threads. 

\subsection{Effect of Abelian point-group symmetry}

In order to quantify the benefits of explicitly considering Abelian point-group symmetry in the computation of CD-CCSD analytic gradients, we compare the theoretical factor of reduction due to symmetry (FRS) with the achieved one for the $\mathcal{O}(O^2V^4)$ scaling contribution in the contraction between the \textit{vvvv} block of the two-body density matrix and the differentiated ERIs for coronene ($D_{2h}$, cc-pVDZ basis set) and azobenzene ($C_{2h}$, aug-cc-pVDZ basis set). Theoretical FRSs are defined as the ratio between the total number of floating-point operations required for the evaluation of a given contraction without exploiting point-group symmetry and when symmetry is enforced. Achieved FRSs are obtained as the ratio between the CPU time necessary for the execution of the same contraction without the use of symmetry and with symmetry. 

For the calculations on coronene, the theoretical FRS for the aforementioned contraction is equal to 56, whereas the achieved FRS is 19. For the azobenzene molecule, on the other hand, the theoretical FRS is equal to 13, while the achieved FRS is 9. The difference between the two values for both systems can be explained by the fact that the explicit inclusion of Abelian point-group symmetry within the implementation makes it necessary to use at least two nested loops over the irreducible representations, leading to a larger number of smaller BLAS matrix-matrix multiplications, impacting the overall efficiency of the code and reducing the formal gain associated with symmetry. It should be noted, however, that both the theoretical and achieved FRSs are comparable to the one reported by Nottoli \textit{et al.}~\cite{Nottoli2023} for the PPL contraction in the solution of the CD-CCSD amplitude equations for similar systems. 

\section{Conclusion and Outlook}

In the present paper we have reported on an efficient implementation of closed-shell CCSD analytic gradients based on the CD of two-electron integrals. The main element of novelty of our implementation is that it fully exploits Abelian point-group symmetry through the DPD scheme, thus noticeably reducing the computational cost of CC analytic gradient calculations and geometry optimizations. Our code makes use of a symmetry-adapted version of the two-step CD algorithm, 
speeding up the decomposition procedure further and, at the same time, yielding symmetry-blocked CVs. The CD formalism allowed us to naturally rewrite the equations for CC first derivatives in terms of three-index intermediates, therefore eliminating the need to store $OV^3$ and $V^4$ quantities and greatly reducing the overall RAM requirements for CCSD gradient calculations. Moreover, as suggested by Aquilante \textit{et al.}~\cite{aquilante2008analytic} and carried out in recent works on CD analytic derivatives~\cite{schnack2022efficient,bostrom2014analytical,delcey2014analytical}, we exploited the formal equivalence between CD and RI/DF to rewrite the ERI derivatives with respect to nuclear displacements in the Cholesky basis and contracted the obtained tensors with CCSD density matrices on-the-fly. In order to integrate the differentiation step with the rest of the (symmetry-adapted) code, integral derivatives were computed with respect to symmetry-adapted displacements, transforming the gradient into a Cartesian coordinate representation only at the end. 

Our implementation was tested on medium-sized systems, consisting of up to 900 basis functions. The computational gains due to symmetry inclusion and parallelization have been verified by computing factors of reduction due to symmetry and plotting speedup graphs for the most expensive contributions in CCSD gradient calculations.

Future work will focus on the extension of our current implementation for open-shell cases, along the lines of Refs.~\cite{gauss1991coupled,gauss1991qrhf,GAUSS1991207rohf}. Other issues of interest are the (perturbative) inclusion of triple excitations as well as the extension of the CD treatment of CC analytic derivatives to higher than first derivatives. All of this will render CD-based CC calculations an important part of the toolbox of high-accuracy computational chemistry.

\section*{Acknowledgement}

This paper is dedicated to the memory of John F. Stanton in recognition of his important contribution to coupled-cluster and equation-of-motion coupled-cluster theory. 
Work in Pisa was supported by the Italian Ministry of Research (PRIN 2022) under grant  2022WZ8LME\_002 and from ICSC-Centro Nazionale di Ricerca in High Performance Computing, Big Data, and Quantum Computing, funded by the European Union-NextGenerationEU-PNRR, Missione 4 Componente 2 Investimento 1.4, while work in Mainz has been supported by the Deutsche Forschungsgemeinschaft (DFG, German Research Foundation) in the framework of the collaborative research center "Multiscale Simulation Methods for Soft-Matter Systems" (TRR 146) under Project No. 233630050.

\printbibliography

\appendix

\section{Intermediates in CD-CCSD Lambda equations}

\label{appendix:lambda}

\begin{align}
    &t^P_{ia} = \sum_et_i^eL^P_{ae}, \\
    &t^P_{ai} = \sum_{m}t_m^aL^P_{mi}, \\
    &t^P_{ij} = \sum_et_i^eL^P_{ej}, \\
    &t^P_{ab} = \sum_mt_m^aL^P_{bm}, \\
    &F_{MI} = \sum_P\left\{\sum_e\left[\sum_{nf}\left(2\Tilde{\tau}_{In}^{Ef} - \Tilde{\tau}_{In}^{Fe}\right)L^P_{nf} - t^P_{EI}\right]L^P_{ME} + 2t_P L^P_{MI}\right\}, \\
    &F_{AE} = -\sum_P\sum_m\left[\sum_{nf}\left(2\Tilde{\tau}_{Mn}^{Af} - \Tilde{\tau}_{Mn}^{Fa}\right)L^P_{nf} + t^P_{ma}\right]L^P_{ME} \notag\\
    &\qquad\quad+ 2\sum_P t_P L^P_{AE}, \\
    &\mathcal{F}_{ME}=2\sum_P t_PL^P_{ME} - \sum_P\sum_n L^P_{NE}t^P_{NM}, \\
    &\mathcal{F}_{MI} = F_{MI} + \sum_e t_I^E\mathcal{F}_{EM},\\
    &\mathcal{F}_{AE} = F_{AE} - \sum_m t_M^A\mathcal{F}_{EM},\\
    &\mathcal{G}_{MI} = \sum_m\sum_{ef}\Tilde{t}^{Ef}_{Mn}\lambda_{Ef}^{In}, \\
    &\mathcal{G}_{AE} = -\sum_{mn}\sum_{f}\Tilde{t}^{Ef}_{Mn}\lambda^{Mn}_{Af}, \\
    &\mathcal{G}_P = \sum_{ai}\mathcal{G}^P_{ai}t^a_i - \sum_{ae}\mathcal{G}_{ae}L^P_{ae}, \\
    &\mathcal{G}^P_{AI} = \sum_e \mathcal{G}_{AE}L^P_{EI}, \\
    &\mathcal{G}^P_{IA} = \sum_m \mathcal{G}_{MI}L^P_{AM}, \\
    &\lambda^P_{IA} = \sum_e\lambda_E^IL^P_{EA}, \\
    &\Bar{\lambda}^P_{IA} = \sum_e\lambda_E^It^P_{EA}, \\
    &\mathcal{V}_{MnIj} = \sum_{Ef} \tau_{Mn}^{Ef} \lambda_{Ef}^{Ij}\\
    &\mathcal{V}^\prime_{IjKl} = \sum_{ef}\tau_{Ij}^{Ef}\Tilde{\lambda}_{Ef}^{Kl}, \\
    &\mathcal{V}_{IjKa} = \sum_e\lambda_{Ea}^{Ij}t_K^E, \\
    &\mathcal{V}_{IjkA} = \sum_e\lambda_{eA}^{Ij}t_k^e,
\end{align}
\begin{align}
    &v^P_{AI} = \sum_{em}\left(\Tilde{\lambda}_{Ae}^{Im}L^P_{em} + \Tilde{\lambda}_{Ae}^{Im}t^P_{me} + 2\mathcal{V}_{AIem}L^P_{em}\right), \\
    &v^P_{IJ} = \sum_{kl}\left(\mathcal{V}^\prime_{IjKl}L^P_{jl} + \mathcal{V}^\prime_{IjKl}t^P_{lk}\right) -\sum_{ef}\mathcal{V}^\prime_{IeJf}L^P_{ef}, \\
    &\mathcal{V}_{AIbj} = \sum_{em}\Tilde{\lambda}_{Ae}^{Im}\left(2t_{Mj}^{Eb} - t_{Jm}^{Eb} - t_m^et_b^j\right), \\
    &\mathcal{V}^\prime_{AIbj} = \mathcal{V}_{AIbj} + \sum_{em}\left(\Tilde{\lambda}_{Ae}^{Im} + \frac{1}{2}\lambda_{Ea}^{Im}\right)\left(t_{Mj}^{Eb} + t_M^Et_j^b\right), \\
    &\mathcal{W}_{MnIj} = \bra{Mn}\ket{Ij} + P_-(ij)\sum_P t^P_{jn}L^P_{MI} + \sum_{ef}\tau_{Ij}^{Ef}\bra{Mn}\ket{Ef}, \\
    &\mathcal{W}_{AbEf} = \sum_P \left[ \left( L_{AE}^P - t_{AE}^P \right) L_{bf}^P - \sum_m t_m^b L_{AE}^P L_{mf}^P \right], \\
    &2\mathcal{W}_{MbEj} + \mathcal{W}_{MbeJ} = 2\bra{Mb}\ket{Ej} - \bra{Mb}\ket{Je}  \\
    &\qquad\quad + \sum_P \left[2\left(t^P_{jb} - t^P_{bj}\right)L^P_{ME} + t^P_{be}L^P_{jm} - t^P_{jm}L^P_{be}\right]\notag\\
    &\qquad\quad + \frac{1}{2}\sum_n\sum_f \left[2t_{Jn}^{Fb} - \left(t_{Jn}^{Fb} + 2t_J^Ft_n^b\right)\right]\left(2\bra{Mn}\ket{Ef} - \bra{Mn}\ket{Fe}\right),\notag\\
    &\mathcal{W}_{MbeJ} = -\bra{Mb}\ket{Je} + \sum_P\left(t^P_{be}L^P_{jm} - t^P_{jm}L^P_{be}\right) \\
    &\qquad+ \frac{1}{2}\sum_n\sum_f \left(t_{Jn}^{Fb} + 2t_J^Ft_n^b\right)\bra{Mn}\ket{Fe},\notag \\
    &\Tilde{\Tilde{\mathcal{W}}}_{MbEj} = \bra{Mb}\ket{Ej} + \frac{1}{2}\sum_{nf}t_{Jn}^{Fb}\bra{Mn}\ket{Ef} , \\
    &\Tilde{\Tilde{\mathcal{W}}}_{MbeJ} = -\bra{Mb}\ket{Je} + \frac{1}{2}\sum_{nf}t_{Jn}^{Fb}\bra{Mn}\ket{Fe}, \\
    &\mathcal{W}_{MnIe} = \sum_P\left(t^P_{IM}+L^P_{IM}\right)L^P_{en}, \\
    &\mathcal{W}_{MbIj} = \bra{Mb}\ket{Ij} - \sum_et_{Ij}^{bE}\mathcal{F}_{ME} - \sum_nt_n^b\mathcal{W}_{MnIj}\\
    &\qquad\quad+\sum_{ef}\tau_{Ij}^{Ef}\bra{Mb}\ket{Ef} + \sum_e\left(t_I^E\Tilde{\Tilde{\mathcal{W}}}_{MbEj} - t_j^e\Tilde{\Tilde{\mathcal{W}}}_{MbeJ}\right) \notag\\
    &\qquad\quad+\frac{1}{2}\sum_{ne}\left(2t_{Jn}^{Be}-t_{Jn}^{Eb}\right)\left(2\bra{Mn}\ket{Ie}-\bra{Nm}\ket{Ie}\right)\notag\\
&\qquad\quad-\sum_{ne}\left(t_{In}^{Eb}\bra{Nm}\ket{Je} + \frac{1}{2}t_{Jn}^{Eb}\bra{Nm}\ket{Ie}\right),\notag\\
    &\Tilde{\mathcal{W}}_{MbEj} = \mathcal{W}_{MbEj} - \bra{Mb}\ket{Ej} - \sum_P\left(t^P_{jb}-t^P_{bj}\right)L^P_{ME}\\
    &\qquad\quad- \sum_{nf}t_J^Ft_n^b\bra{Mn}\ket{Ef},\notag\\
    &\Tilde{\mathcal{W}}_{MbeJ} = \mathcal{W}_{MbeJ} - \bra{Mb}\ket{Je} - \sum_P\left(t^P_{be}L^P_{jm}-t^P_{jm}L^P_{be}\right)\\
    &\qquad\quad-\sum_{nf}t_J^Ft_n^b\bra{Mn}\ket{Fe}\notag.
\end{align}

\section{CCSD one-body density matrices}

\label{appendix:d1cc}

\begin{align}
  & D_{IJ} = -\frac{1}{2} P_+(ij) \sum_{m} \sum_{ef} \tilde{t}_{im}^{ef} \lambda_{Ef}^{Jm} -\frac{1}{2} P_+(ij) \sum_{e} t_I^E \lambda_E^J,\\
  & D_{AB} = \frac{1}{2} P_+(ab) \sum_{mn} \sum_{e} \tilde{t}_{mn}^{ae} \lambda_{Be}^{Mn} + \frac{1}{2} P_+(ab) \sum_{m} t_M^A \lambda_B^M,\\
  & D_{AI} = \frac{1}{2} \left[ t_I^A + \lambda_A^I + \sum_{m} \sum_{e} (\tilde{t}_{im}^{ae} - t_{I}^E t_M^A) \lambda_E^M - \sum_m \mathcal{G}_{IM} t_M^A + \sum_e \mathcal{G}_{EA} t_I^E \right],
\end{align}
where $P_+(pq)$ is the symmetric permutation operator:
\begin{equation}
    P_+(pq) = 1 + \mathcal{P}(pq).
\end{equation}

\section{CCSD two-body density matrices}

\label{appendix:d2cc}

\begin{align}
  &\Gamma_{AbCd} = \frac{1}{2} P_+(ab,cd) \mathcal{V}_{AbCd},\\
  &\Gamma_{IjKl} = \frac{1}{2} P_+(ij,kl) \mathcal{V}_{IjKl},\\
  &\Gamma_{IbJa} = P_+(ia,jb) \mathcal{V}_{IbJa} - \frac{1}{2}P_+(ia,jb) \sum_{m} \sum_e t_I^E t_m^a \lambda_{Eb}^{Jm},\\
  &\Gamma_{IbAj} = -P_+(ia,jb) \mathcal{V}_{IbjA} - \frac{1}{2}P_+(ia,jb) \sum_{m} \sum_e t_I^E t_M^A \lambda_{Eb}^{Mj} + \frac{1}{2}P_+(ia,jb) t_I^A \lambda_b^j,\\
&\Gamma_{AbIj} = \frac{1}{2} \tau_{Ij}^{Ab} + \frac{1}{2} \lambda_{Ab}^{Ij} + \frac{1}{2} \sum_{mn} \tau_{Mn}^{Ab} \mathcal{V}_{IjMn} - \frac{1}{2}P_-(ij) \sum_m \tau_{Mj}^{Ab} \left(\mathcal{G}_{IM} + \sum_e \lambda_E^M t_I^E \right)\\
  &\qquad + \frac{1}{2} P_-(ab) \sum_e \tau_{Ij}^{Eb} \left(\mathcal{G}_{EA} - \sum_m \lambda_E^M t_M^A \right) + \frac{1}{2} P_+(ia,jb) \bigg[\sum_{m} \sum_e (t_{Mj}^{Ae} + 2t_M^A t_j^e) \mathcal{V}_{IeMb} \notag\\
  &\qquad - \sum_{m} \sum_e (\tilde{t}_{im}^{ae} - 2t_M^At_I^E)(\mathcal{V}_{JemB} + \lambda_E^M t_j^B) - \sum_{m} \sum_e t_{Im}^{Ae} \mathcal{V}_{JeMb} \bigg] + \notag\\
  &\qquad + \frac{3}{2} P_+(ia,jb) \sum_{m} \sum_e t_m^b t_j^e \lambda_e^m t_I^A,\notag\\
  &\label{eq:tpdm_vvvo} \Gamma_{AbCi} = \frac{1}{2} \sum_m \lambda_C^M \tau_{Mi}^{Ab} + \frac{1}{2} \sum_m t_M^C \lambda_{Ab}^{Mi} - \frac{1}{2} \sum_e \mathcal{V}_{CeAb} t_i^e + \sum_m \mathcal{V}_{MaIc} t_M^B \\
  &\qquad - \sum_m \mathcal{V}_{MbiC} t_M^A - \frac{1}{2} \mathcal{G}_{CA} t_i^b, \notag\\
  &\Gamma_{IjKa} = -\frac{1}{2} \sum_e \lambda_E^K \tau_{Ij}^{Ea} -\frac{1}{2} \sum_e t_K^E \lambda_{Ea}^{Ij} + \frac{1}{2} \sum_m \mathcal{V}_{IjKm} t_m^a + \sum_f \mathcal{V}_{KaIf} t_j^f \\
  &\qquad - \sum_f \mathcal{V}_{KajF} t_I^F  - \frac{1}{2} \mathcal{G}_{IK} t_j^a,\notag
\end{align}
where:
\begin{align}
  &\mathcal{V}_{AbEf} = \sum_{mn} \tau_{Mn}^{Ef} \lambda_{Ab}^{Mn},\\
  &\mathcal{V}_{IbJa} = -\frac{1}{2} \sum_{m} \sum_e t_{Jm}^{Eb} \lambda_{Ea}^{Im},\\
  &\mathcal{V}_{IbjA} = -\frac{1}{2} \sum_{m} \sum_e t_{Be}^{Jm} \lambda_{AE}^{IM} - \frac{1}{2} \sum_{m} \sum_e t_{jm}^{be} \lambda_{Ae}^{Im}.
\end{align}

\end{document}